\documentclass[preprint]{common/llncs}
\pdfoutput=1
\usepackage{times}
\usepackage{amscd}
\usepackage{amsmath}
\usepackage{url}
\usepackage{graphicx}
\usepackage{lscape}
\usepackage{listings}
\lstMakeShortInline[language=C]!

\usepackage{times}
\SetMathAlphabet{\mathtt}{normal}{OT1}{pcr}{m}{n}
\SetMathAlphabet{\mathtt}{bold}{OT1}{pcr}{bx}{n}

\usepackage{amsmath}
\usepackage{amssymb} 

\newcommand{\CUT}[1]{}

\newcommand{\secref}[1]{Section~\ref{#1}}
\newcommand{\tblref}[1]{Table~\ref{#1}}
\newcommand{\figref}[1]{Figure~\ref{#1}}

\providecommand{\bftt}[1]{{\ttfamily\bfseries{}#1}}

\providecommand{\kw}[1]{\bftt{#1}}
\providecommand{\nt}[1]{{\rmfamily\itshape{#1}}}

\newcount\timeHH
\newcount\timeMM
\timeHH=\time
\divide\timeHH by 60
\timeMM=\time
\multiply\count255 by -60 \advance\timeMM by \count255
\newcommand{\timestamp}{%
  \today{} ---
  \ifnum\timeHH<10 0\fi\number\timeHH\,:\,\ifnum\timeMM<10 0\fi\number\timeMM}

\usepackage{common/code}

\title{Measuring NUMA effects with the STREAM benchmark}
\titlerunning{Measuring NUMA effects with the STREAM benchmark}
\author{Lars Bergstrom}
\authorrunning{Lars Bergstrom}
\institute{University of Chicago, Chicago IL 60637, USA\\
\email{larsberg@cs.uchicago.edu}}

\date{\today}

\begin{document}

\maketitle
\thispagestyle{empty}

\sloppy

%
\begin{abstract}
Modern high-end machines feature multiple processor packages, each of which
contains multiple independent cores and integrated memory controllers connected
directly to dedicated physical RAM.
These packages are connected via a shared bus, creating a system with a
heterogeneous memory hierarchy.
Since this shared bus has less bandwidth than the sum of the links to memory,
aggregate memory bandwidth is higher when parallel threads all access memory
local to their processor package than when they access memory attached to a
remote package.

But, the impact of this heterogeneous memory architecture is not easily
understood from vendor benchmarks.
Even where these measurements are available, they provide only best-case memory
throughput. 
This work presents a series of modifications to the well-known STREAM benchmark
to measure the effects of NUMA on both a 48-core AMD Opteron machine and a
32-core Intel Xeon machine.
\end{abstract}



\section{Introduction}
Inexpensive multicore processors and accessible multiprocessor motherboards
have brought all of the challenges inherent in parallel programming with large numbers of
threads with non-uniform memory access (NUMA) into the foreground.
Functional programming languages are a particularly interesting approach
to programming parallel systems, since they provide a high-level programming
model that avoids many of the pitfalls of imperative parallel programming.
But while functional languages may seem like a better fit for parallelism due to
their ability to compute independently while avoiding race conditions and
locality issues with shared memory mutation, implementing a scalable functional
parallel programming language is still challenging.
Since functional languages are value-oriented, their performance is highly
dependent upon their memory system.
This system is often the major limiting to improved performance in these
systems~\cite{multicore-haskell,intel-private-heap}.

Our group has been working on the design and implementation of a
parallel functional language to address the opportunity afforded by
multicore processors.
In this paper, we describe some benchmarks we have used to measure top-end
AMD and Intel machines to assist in the design and tuning of our parallel
garbage collector.

This paper makes the following contributions:
\begin{enumerate}
  \item
    We describe the architecture of and concretely measure the bandwidth and
    latency due to the memory topology in both a 48-core AMD Opteron server and
    a 32-core Intel Xeon server.
    Looking only at technical documents, it is difficult to understand how much
    bandwidth is achievable from realistic programs and how the latency of
    memory access changes with increased bus saturation.
\end{enumerate}%




\subsection{AMD Hardware}
\label{intro:amdhardware}
\begin{figure}
\centering
\includegraphics[scale=0.7]{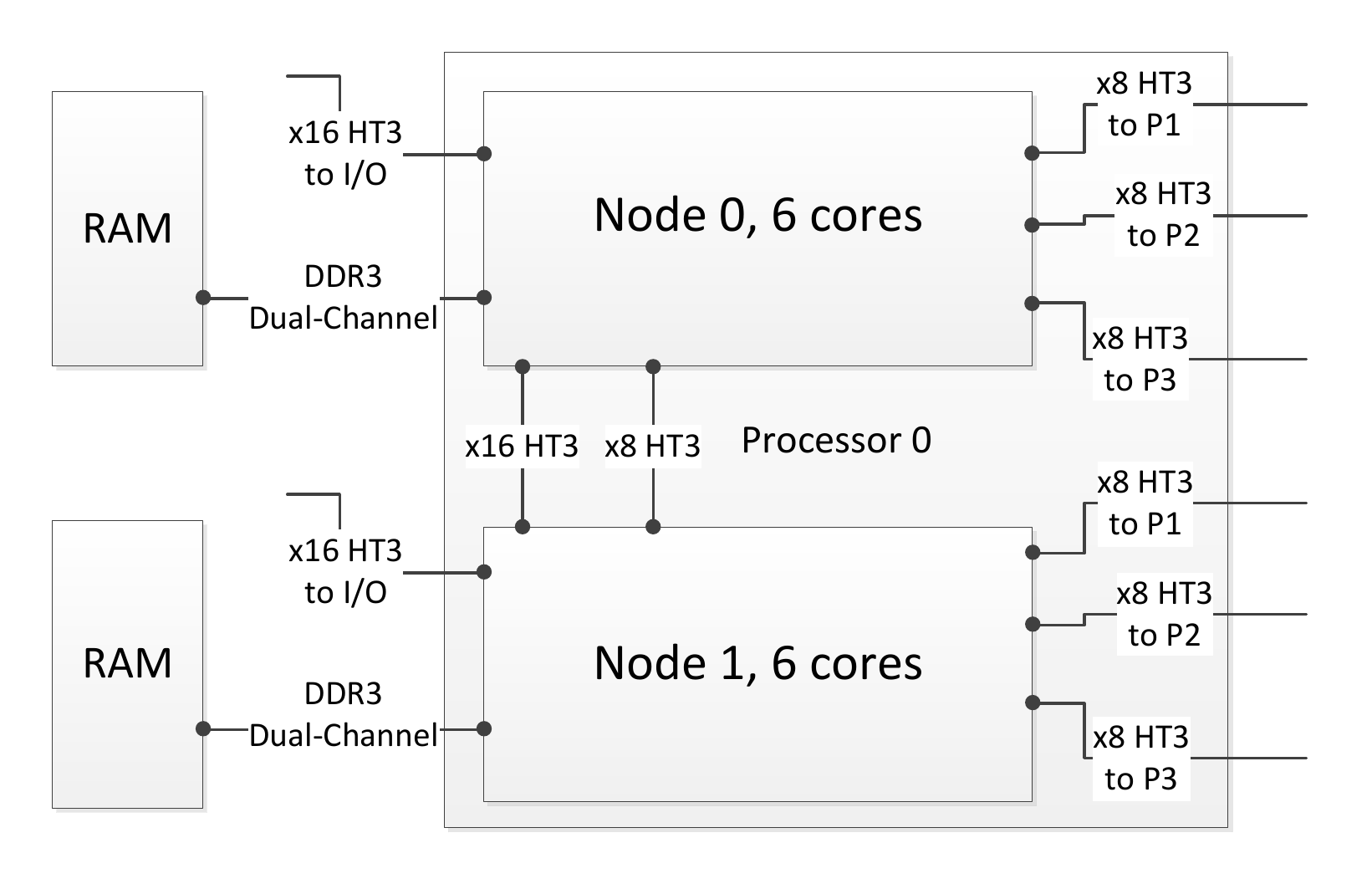}
\caption{
Interconnects for one processor in a quad AMD Opteron machine.
}
\label{amd:magny}
\end{figure}

Our AMD benchmark machine is a Dell PowerEdge R815 server, outfitted
with 48~cores and 128~GB physical memory.
The 48~cores are provided by four AMD Opteron 6172 ``Magny Cours'' processors~\cite{magny-cours,opteron},
each of which fits into a single G34 socket.
Each processor contains two nodes, and each node has six cores.
The 128~GB physical memory is provided by thirty-two 4~GB dual ranked RDIMMs,
evenly distributed among four sets of eight sockets, with one set for each processor.
As shown in \figref{amd:magny}, these nodes, processors, and RAM chips form a
hierarchy with significant differences in available memory bandwidth and number
of hops required, depending upon the source processor core and the target physical memory location.
Each 6~core node (die) has a dual-channel double data rate 3 (DDR3) memory
configuration running at 1333~MHz from its private memory controller to its own
memory bank.
There are two of these nodes in each processor package. 
This processor topology is also laid out in \tblref{amd:topology}.

\begin{table}
  \begin{center}
  \begin{tabular}{r | c | c}
Component & Hierarchy & \# Total\\
\hline
Processor & 4 per machine & 4\\
Node & 2 per processor & 8\\
Core & 6 per node & 48 \\
\end{tabular}
\end{center}
\caption{
  Processor topology of the AMD machine.
}
\label{amd:topology}
\end{table}%

Bandwidth between each of the nodes and I/O devices is provided by four 16-bit
HyperTransport 3 (HT3) ports, which can each be separated into two 8-bit HT3
links.
Each 8-bit HT3 link has 6.4~GB/s of bandwidth. 
The two nodes within a package are configured with a full 16-bit link and an
extra 8-bit link connecting them. 
Three 8-bit links connect each node to the other three packages in this four
package configuration. 
The remaining 16-bit link is used for I/O.
\figref{numa:amdbandwidth} shows the bandwidth available between the different
elements in the hierarchy.
\begin{table}
  \begin{center}
  \begin{tabular}{r | c }
  \multicolumn{1}{c|}{} & Bandwidth (GB/s)\\
\hline
Local Memory & 21.3 \\
Node in same package & 19.2 \\
Node on another package & 6.4 \\
\end{tabular}
\end{center}
\caption{
  Theoretical bandwidth available between a single node (6~cores) and the rest of an AMD Opteron 4P system.
}
\label{numa:amdbandwidth}
\end{table}%

Each core operates at 2.1~GHz and has 64~KB each of instruction and data L1 cache and 512~KB of L2 cache.
Each node has 6~MB of L3 cache physically present, but, by default, 1~MB is
reserved to speed up cross-node cache probes.





\subsection{Intel Hardware}
\label{intro:intelhardware}
\begin{figure}
\centering
\includegraphics[scale=0.7]{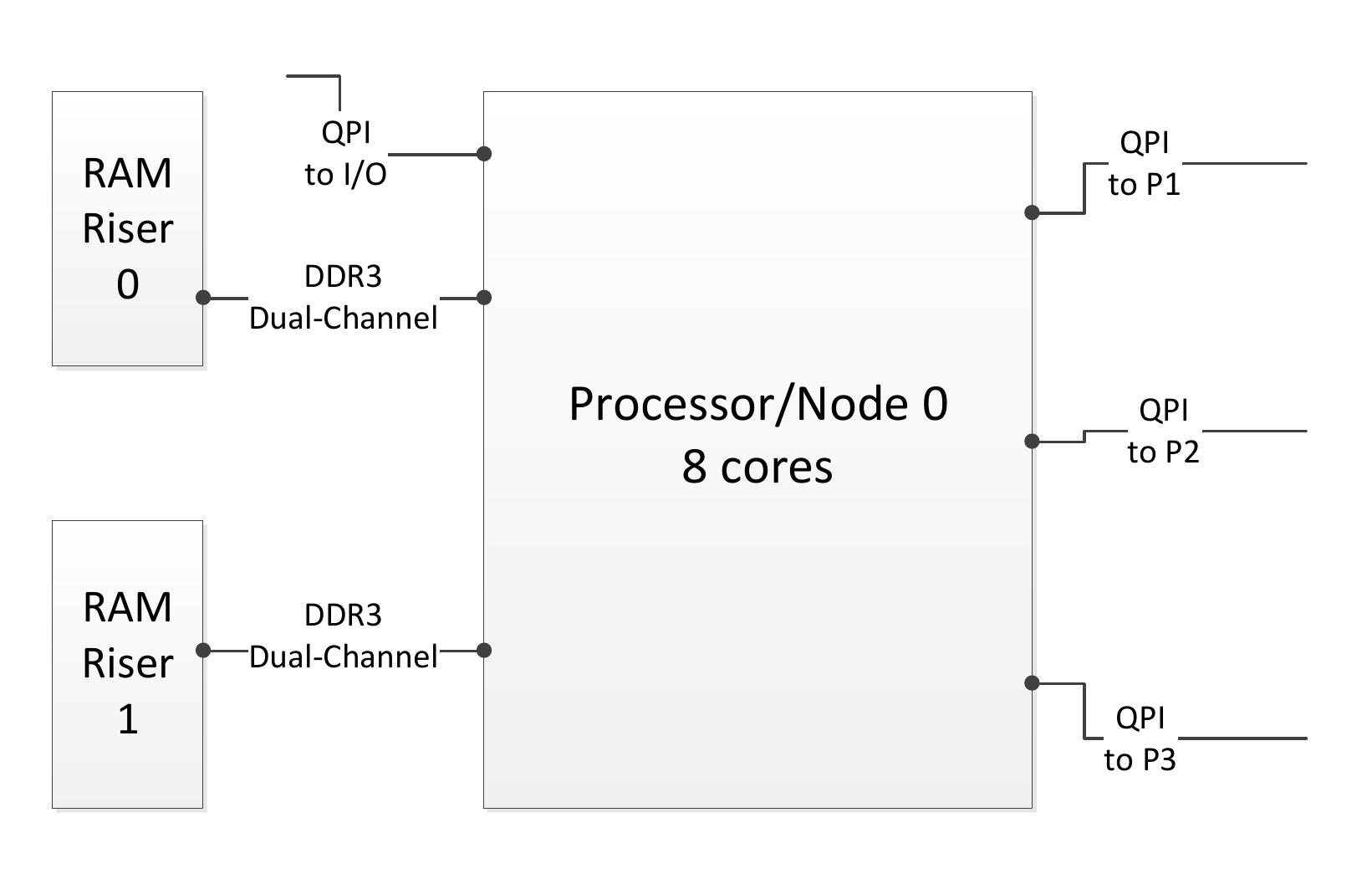}
\caption{
Interconnects for one processor in a quad Intel Xeon machine.
}
\label{intel:xeon}
\end{figure}

The Intel benchmark machine is a QSSC-S4R server with 32~cores and 256~GB
physical memory.
The 32~cores are provided by four Intel Xeon X7560 processors~\cite{xeon,qssc}.
Each processor contains 8~cores, which can be but are not configured to run with
2~simultaneous multithreads (SMT).
This topology is laid out in \tblref{intel:topology}.
As shown in \figref{intel:xeon}, these nodes, processors, and RAM chips form a
hierarchy, but this hierarchy is more uniform than that of the AMD machine.

\begin{table}
  \begin{center}
  \begin{tabular}{r | c | c}
Component & Hierarchy & \# Total\\
\hline
Processor & 4 per machine & 4\\
Node & 1 per processor & 4\\
Core & 8 per node & 32\\
\end{tabular}
\end{center}
\caption{
  Processor topology of the Intel machine.
}
\label{intel:topology}
\end{table}%
Each of the nodes is connected to two memory risers, each of which has a
dual-channel DDR3 1066~MHz connection.
The 4~nodes are fully connected by full-width Intel QuickPath Interconnect (QPI)
links.
\figref{numa:intelbandwidth} shows the bandwidth available between the different
elements in the hierarchy.

\begin{table}
  \begin{center}
  \begin{tabular}{r | c }
  \multicolumn{1}{c|}{} & Bandwidth (GB/s)\\
\hline
Local Memory & 17.1 \\
Other Node & 25.6 \\
\end{tabular}
\end{center}
\caption{
  Theoretical bandwidth available between a single node (8~cores) and the rest of an Intel Xeon system.
}
\label{numa:intelbandwidth}
\end{table}%
Each core operates at 2.266~GHz and 32~KB each of instruction and data L1 cache
and 256~KB of L2 cache.
Each node has 24~MB of L3 cache physically present but, by default, 3~MB is
reserved to speed up both cross-node and cross-core caching.



\section{Measuring NUMA effects}
\label{sec:numa}

In \secref{intro:amdhardware}, we described the exact hardware configuration and 
memory topology of our 48~core AMD Opteron system.
We also described the 32~core Intel Xeon system in
\secref{intro:intelhardware}. 
These systems are the subject of the NUMA tests below.

\subsection{STREAM benchmark}
The C language STREAM benchmark~\cite{stream} consists of the four operations
listed in \tblref{numa:stream}. 
These synthetic memory bandwidth tests were originally selected to measure
throughput rates for a set of common operations that had significantly different
performance characteristics on vector machines of the time.
On modern hardware, each of these tests achieve similar bandwidth, as memory is
the primary constraint, not floating-point execution.
The \kw{COPY} test, in particular, is representative of the type of work
performed by a copying garbage collector.

\begin{table}
\begin{center}
\begin{tabular}{r | l}
Name & Code\\
\hline
\kw{COPY} & \lstinline!a[i] = b[i];! \\
\kw{SCALE} & \lstinline!a[i] = s*b[i];! \\
\kw{SUM} & \lstinline!a[i] = b[i]+c[i];! \\
\kw{TRIAD} & \lstinline!a[i] = b[i]+s*c[i];! \\
\end{tabular}
\end{center}
\caption{
Basic operations in the STREAM benchmark.
}
\label{numa:stream}
\end{table}%
The existing STREAM benchmark does not support NUMA awareness for either the
location of the running code or the location of the allocated memory. 
We modified the STREAM benchmark to measure the achievable memory bandwidth for
these operations across several allocation and access configurations.
The baseline STREAM benchmark allocated a large, static vector of \kw{double}
values.\footnote{There is no difference in bandwidth when using \kw{long}
  values.}
Our modifications use pthreads and libnuma to control the number and placement
of each piece of running code and corresponding
memory~\cite{butenhof:pthreads-book,libnuma}.

While the STREAM benchmark's suggested array sizes for each processor are larger
than the L3 cache, the tests do not take into account cache block sizes.
We extended the tests with support for strided accesses to provide a measure of
RAM bandwidth in the case of frequent cache misses.
This strided access support also allows us to measure the latency of memory
access.

\subsection{Bandwidth evaluation}
\begin{figure}
\centering
AMD Bandwidth\\
\includegraphics[scale=0.9]{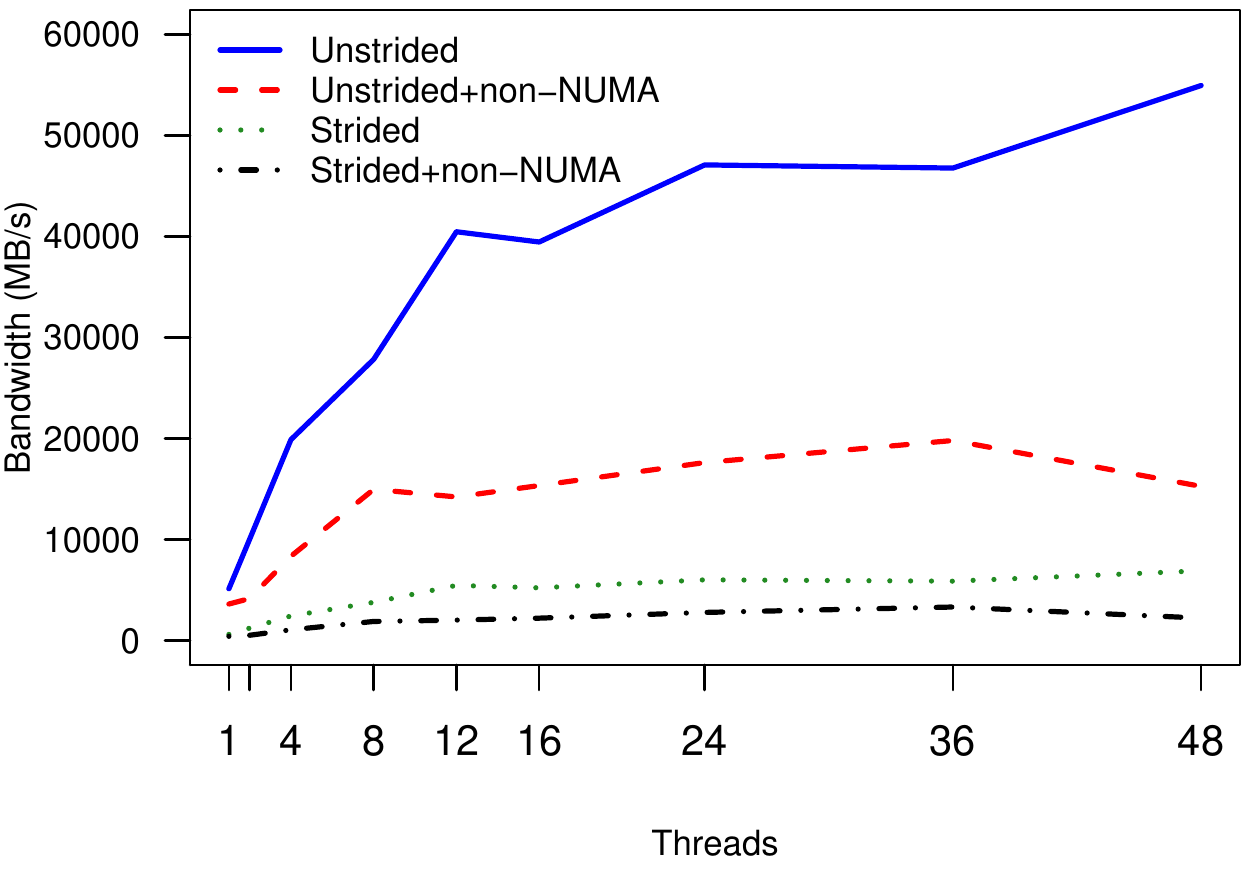} \\[3em]
Intel Bandwidth\\
\includegraphics[scale=0.9]{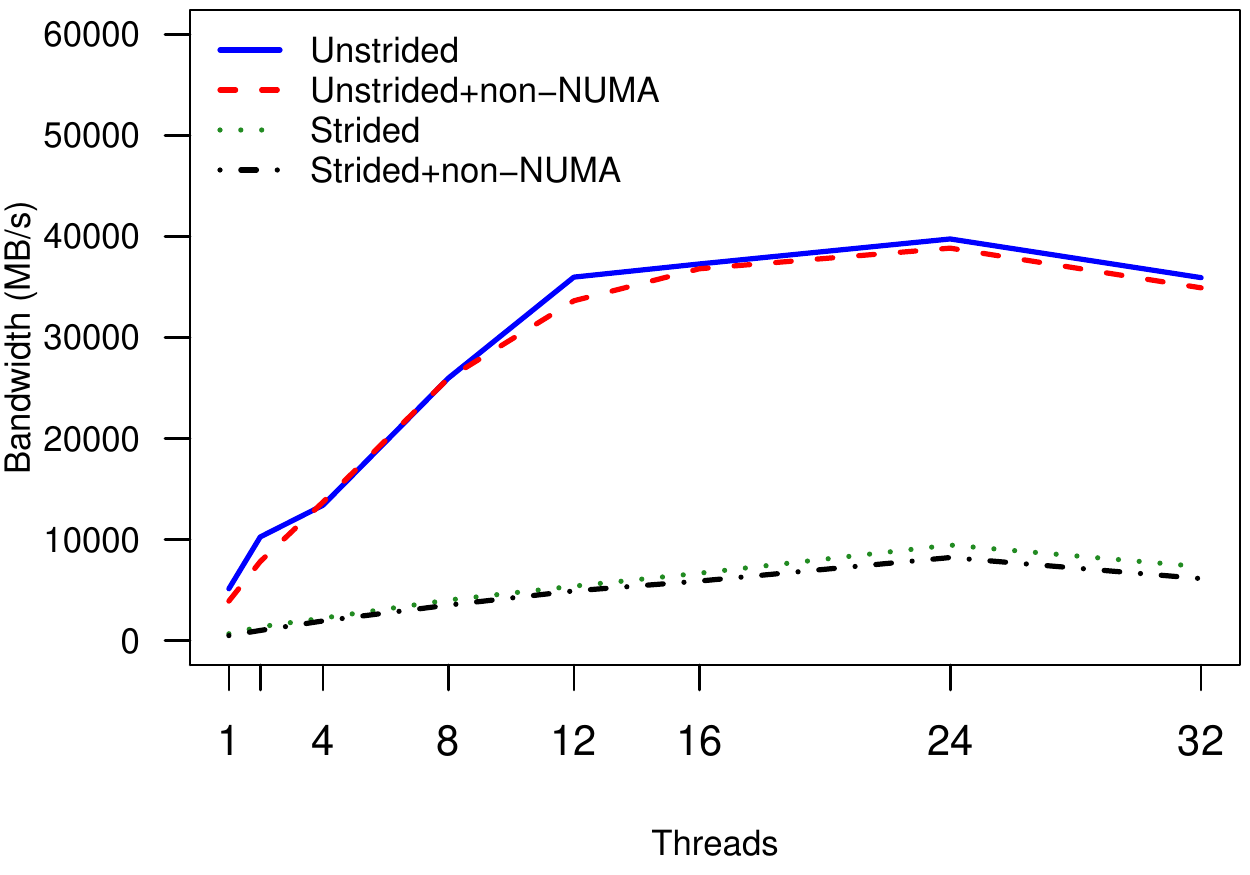}
\caption{
Bandwidth vs. number of threads on the STREAM benchmark, comparing strided and NUMA configurations.
Larger bandwidth is better.
}
\label{numa:evalBandwidth}
\end{figure}

\figref{numa:evalBandwidth} plots the bandwidth, in MB/s, versus the number of
threads.
Larger bandwidth is better.
The results are for the \kw{COPY} test, but all of the tests were within a small
factor.
Four variants of the STREAM benchmarks were used: 
\begin{enumerate}
\item \emph{Unstrided} accesses memory attached to its own node and uses the
  baseline STREAM strategy of sequential access through the array.
\item \emph{Unstrided+non-NUMA} accesses the array sequentially, but is
  guaranteed to access that memory on another package.
\item \emph{Strided} also accesses local memory, but ensures that each access is
  to a new cache block.
\item \emph{Strided+non-NUMA} strides accesses, and also references memory from
  another package.
\end{enumerate}
NUMA aware versions ensure that accessed memory is allocated on the same node as
the thread of execution and that the thread is pinned to the node, using the
libnuma library~\cite{libnuma}.
To do this, the modified benchmark pins the thread to a particular node and then
uses the libnuma allocation API to guarantee that the memory is allocated on the
same node.
The non-NUMA aware versions also pin each thread to a particular node, but then
explicitly allocate memory using libnuma from an entirely separate package (not
just a separate node on the same package).
When there are less threads than cores, we pin threads to new nodes rather than
densely packing a single node.

It should not be surprising that the unstrided variants exhibit roughly eight
times the bandwidth of their strided versions, as cache blocks on these machines 
are 64~bytes and the \kw{double} values accessed are each 8~bytes.
In the NUMA aware cases, scaling continues almost linearly until eight threads
and increases until the maximum number of available cores on both machines.
On AMD hardware, non-NUMA aware code pays a significant penalty and begins to
lose bandwidth where NUMA aware code does not at 48~cores.
On the Intel hardware the gap between NUMA and non-NUMA aware code is very small
even when the number of threads is the same as the number of cores.
But, the Intel hardware does not offer as much peak usable bandwidth for
NUMA-aware code, peaking near 40,000~MB/s whereas we achieve nearly 55,000~MB/s
on the AMD hardware.

\begin{figure}
\centering
AMD Bandwidth\\
\includegraphics[scale=0.9]{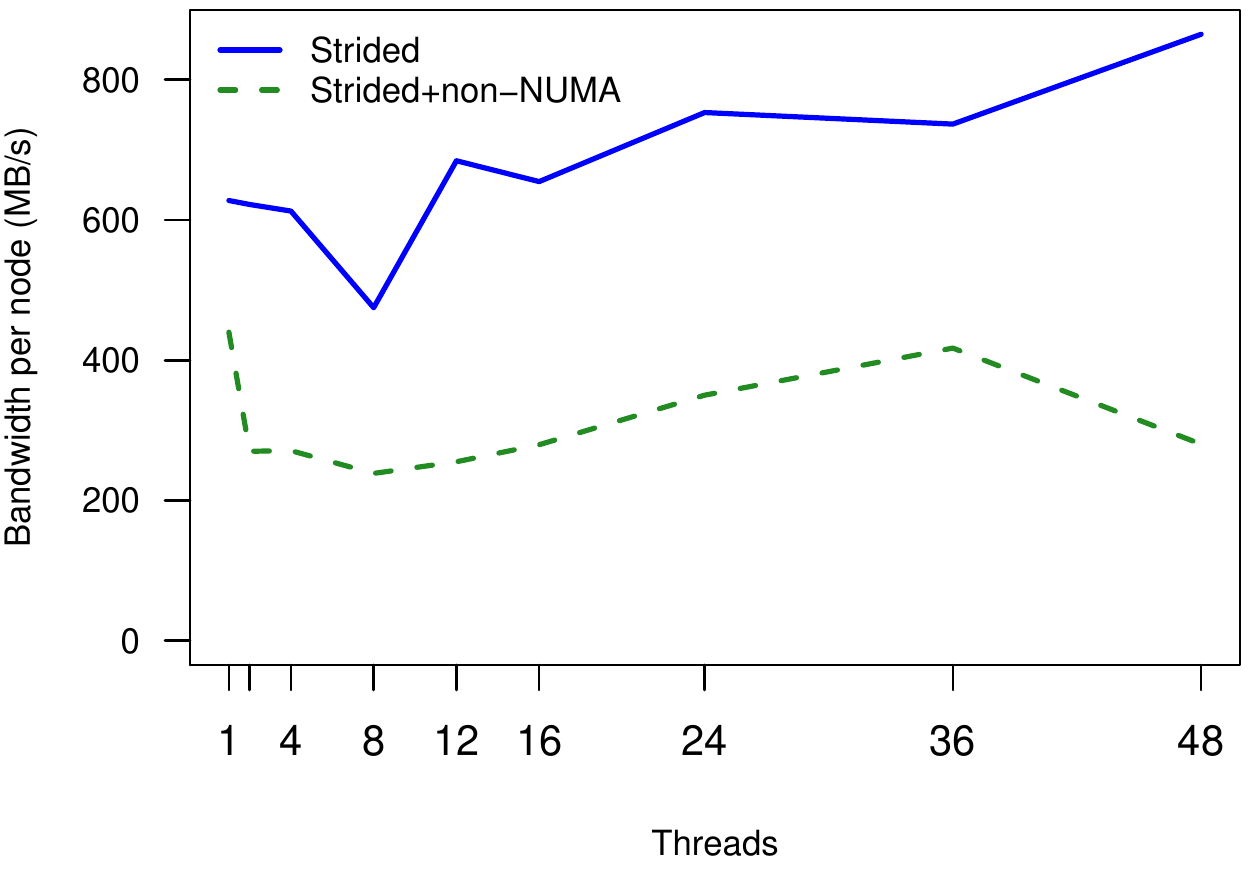} \\[3em]
Intel Bandwidth\\
\includegraphics[scale=0.9]{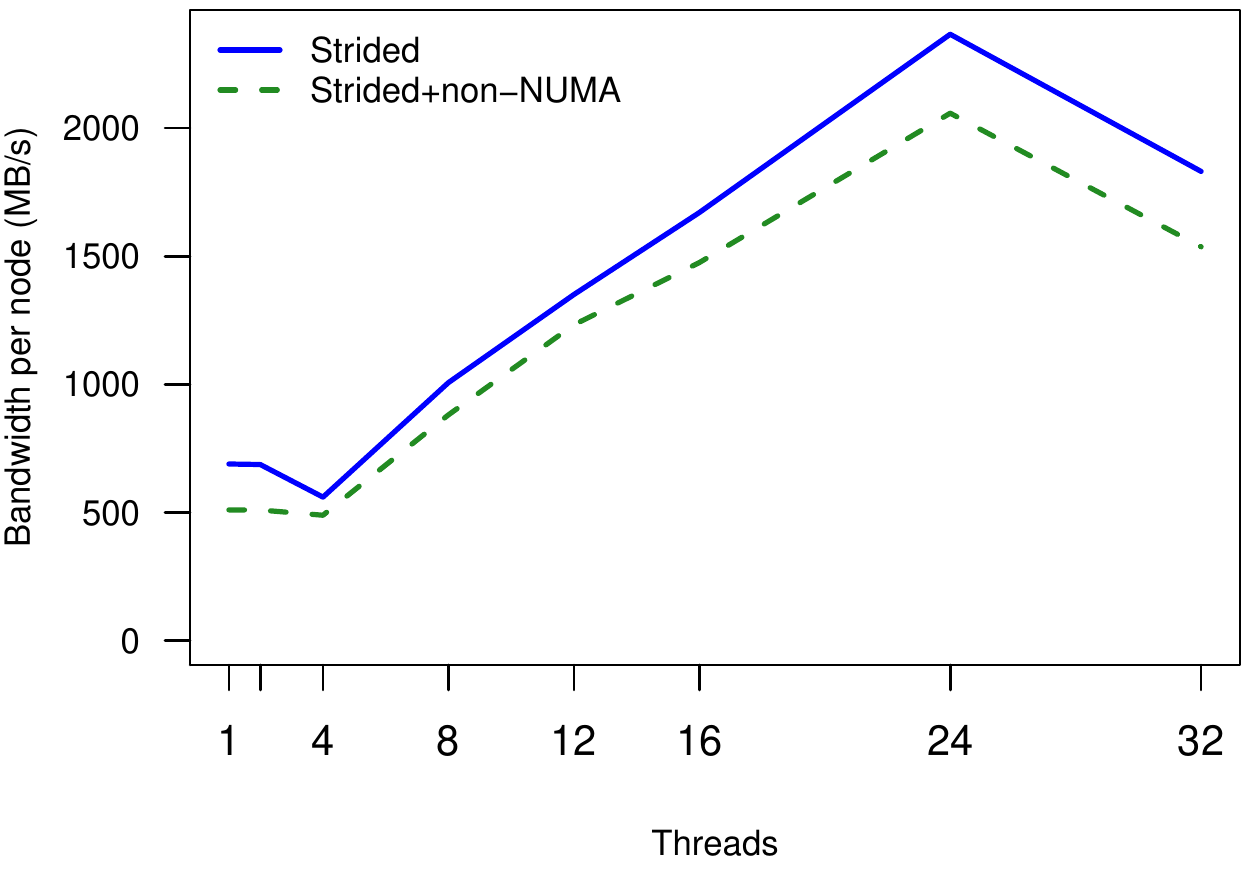}
\caption{
Bandwidth per node vs. number of threads on the STREAM benchmark, comparing
strided configurations.
Larger bandwidth is better.
}
\label{numa:evalBandwidthByNode}
\end{figure}%

\figref{numa:evalBandwidthByNode} plots the bandwidth against threads again,
but this time divided by the number of active nodes to provide a usage data
relative to the theoretical interconnect bandwidth detailed in
\tblref{numa:amdbandwidth} for the AMD machine and \tblref{numa:intelbandwidth}
for the Intel machine.
Our benchmarks allocate threads sparsely on the nodes. 
Therefore, when there are less than 8~threads on the AMD machine, that is also
the number of active nodes.
On the Intel machine with 4~nodes, when there are less than 4~threads, that is
the number of active nodes.
These graphs show that on both machines there is a significant gap between
the theoretical bandwidth and that achieved by the strided \kw{COPY} stream
benchmark.
It is also clear that there is a significant non-NUMA awareness penalty on the
AMD machine but that penalty is less on the Intel machine.
However, the Intel machine begins to reach saturation at 32~cores, whereas the
NUMA aware AMD machine continues to increase per-node bandwidth up to 48~cores. 

\begin{figure}
\centering
AMD Latencies\\
\includegraphics[scale=0.9]{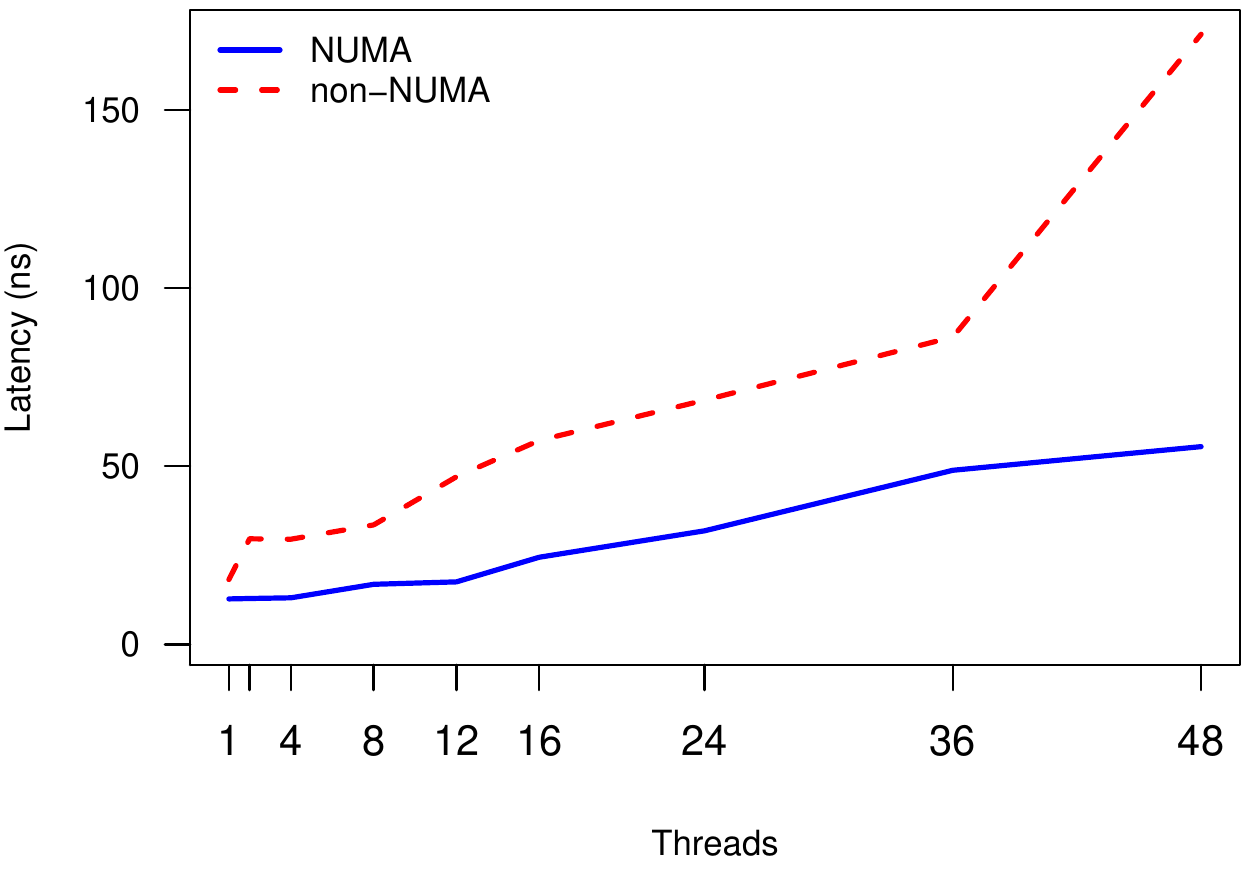} \\[3em]
Intel Latencies\\
\includegraphics[scale=0.9]{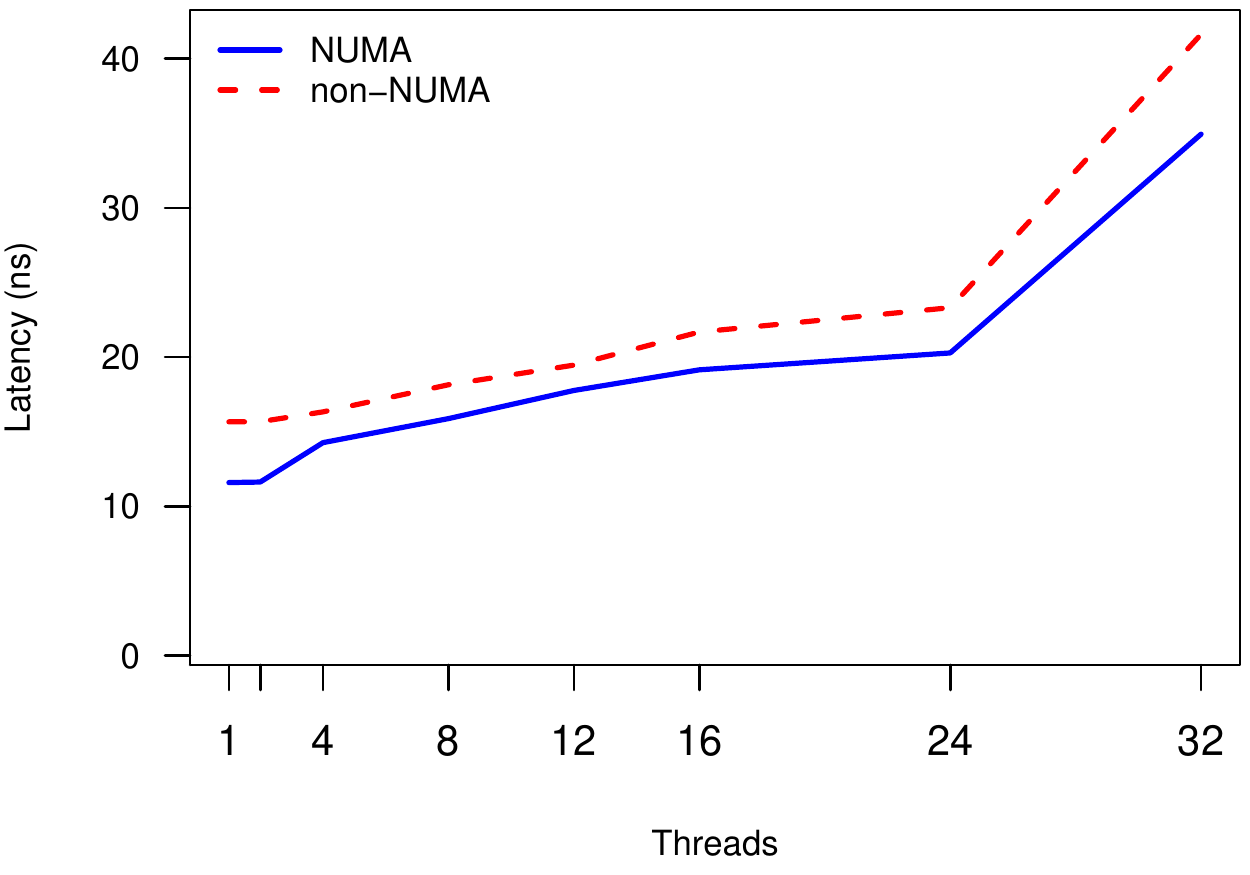}
\caption{
Latency times versus the number of threads on the STREAM benchmark, comparing strided configurations.
Smaller latency times are better.
}
\label{numa:evalLatency}
\end{figure}%

\subsection{Latency evaluation}
\figref{numa:evalLatency} plots the latency times, in nanoseconds, versus the
number of threads.
Smaller latency times are better.
To measure the latency times, we only consider the strided STREAM benchmarks 
to ensure that we are measuring only the time to access RAM, and not the time
to access cache.
As was the case with bandwidth, the AMD machine's NUMA aware tests maintain good
values up to large numbers of processors.
The non-NUMA aware AMD benchmark begins to exhibit high latencies at moderate
numbers of threads.
On the Intel machine, latency numbers remain low until more than 24~cores are in
use, and then the latencies grow similarly for both NUMA aware and non-NUMA
aware code.

On the AMD machine, these benchmarks and evaluations clearly indicate that at
high numbers of threads of executions, poor choices of memory location or code
execution can have significant negative impact on the latency of memory access
and memory bandwidth.
For the Intel machine, memory performance uniformly increases until high numbers
of threads, at which point is uniformly decreases, seeing little effect from
NUMA awareness.
But, none of these effects show up in practice until more than 24~threads are in
use.


%
\section{Conclusion}
\label{sec:concl}
We have described and measured the memory topology of two different high-end
machines using Intel and AMD processors.
These measurements demonstrate that NUMA effects exist and require engineering
beyond that normally employed to achieve good locality and cache use.

Further, we have shown that the NUMA penalty is significantly lower on Intel
systems due to the larger cross-processor bandwidth provided by QPI.
But, the AMD system provides greater total bandwidth for NUMA-aware
applications.

\paragraph{Acknowledgments}
Thanks to Bradford Beckmann for reviewing the breakdown of the AMD G34 socket.
The AMD machine used for the benchmarks was supported by National Science
Foundation Grant CCF-1010568 and this work is additionally supported in part by
National Science Foundation Grant CCF-0811389. 
The views and conclusions contained herein are those of the authors and should
not be interpreted as necessarily representing the official policies or
endorsements, either expressed or implied, of these organizations or the
U.S.\ Government.

Access to the Intel machine was provided by Intel Research.
Thanks to the management, staff, and facilities of the Intel Manycore Testing
Lab.\footnote{Manycore Testing Lab Home:\\
  \url{http://www.intel.com/software/manycoretestinglab}\\
Intel Software Network:\\
\url{http://www.intel.com/software}}

\bibliographystyle{common/alpha}
\bibliography{common/strings-short,common/manticore}

\end{document}